\documentclass{article}

\usepackage{amsmath, amssymb}
\usepackage{physics}

\usepackage{graphicx}
\usepackage{tikz}
\usetikzlibrary{positioning, arrows.meta}

\usepackage{algorithm}
\usepackage{algpseudocode}

\usepackage{listings}
\usepackage{xcolor}

\usepackage{cite}

\usepackage{authblk}

\usepackage{hyperref}

\title{Optimal Approximation of Single Qubit Rotations within a Quantum Circuit}

\author{Gilad Kishony}
\author{Avi Elazari}
\author{Ron Cohen}
\author{Lior Gazit}

\affil{Classiq Technologies, 3 Daniel Frisch Street, Tel Aviv-Yafo, 6473104, Israel}

\date{}

\begin{document}

\maketitle

\begin{abstract}
Fault-tolerant quantum computing typically requires the transpilation of arbitrary quantum circuits into a finite, universal gate set, such as Clifford+T. As a baseline, Diagonal approximation can be used for synthesizing single-qubit Pauli rotations, yielding an approximating sequence with $T$-count that equals $3 \log_2(1/\epsilon)$ for a target precision $\epsilon$. Magnitude Approximation can reduce the $T$-count to only $1 \log_2(1/\epsilon)$ by allowing large residual errors, which are rotations about orthogonal axes. Within a complete quantum circuit, these residual errors can then be absorbed into neighboring gates before they are approximated themselves. Determining the optimal allocation of approximation strategies within a large, multi-qubit circuit presents a significant combinatorial challenge. In this work, we present a linear-time algorithm that guarantees an optimal solution to this problem. We demonstrate that the issue of delegating Magnitude versus Diagonal approximation across a circuit maps formally to a classical 1D Ising model with a spatially varying field. By minimizing the energy of this Hamiltonian, we identify the optimal approximation configuration for each rotation without exponential overhead. Benchmarking our method against standard diagonal approximation on random quantum circuits, we observe an average reduction of 26\% in the total approximating circuit gate count, offering a significant efficiency gain for the implementation of quantum algorithms on near-term and fault-tolerant architectures.

\end{abstract}

\section{Introduction}

Quantum circuit optimization is a critical challenge in the realization of practical quantum computing, particularly for fault-tolerant architectures. Due to hardware constraints and the requirements of Quantum Error Correction (QEC) schemes, quantum computers typically operate using a universal but discrete gate set, such as the Clifford+T set. Consequently, arbitrary unitary operations, specifically single-qubit rotations, cannot be implemented natively and must be approximated using sequences of gates from this finite set.

The efficiency of these approximations directly impacts the execution time and fidelity of quantum algorithms. While the Solovay-Kitaev algorithm provides a general solution with polylogarithmic overhead, it is often suboptimal compared to number-theoretic approaches tailored explicitly for the Clifford+T set. Specifically, for a query unitary which is a single-qubit Pauli rotation, \textit{diagonal approximation} achieves an approximation error of $\epsilon$ with a sequence of $T$-count $3 \log_2(1/\epsilon)$. Since a general unitary $U$ can always be exactly decomposed into a sequence of three rotations (e.g., $X-Z-X$ Euler angles) to be synthesized, the cost of a standard diagonal approximation for such a unitary becomes $9 \log_2(1/\epsilon)$ for a target precision $\epsilon$.

To reduce this overhead, alternative approaches have been explored. These include \textit{Projective Rotations} utilizing fallback approximations \cite{PhysRevLett.114.080502, Kliuchnikov_2023} and \textit{Mixed Approximations} that utilize non-unitary circuits or probabilistic mixtures \cite{PhysRevA.95.022316, hastings2016turninggatesynthesiserrors}. In this work, we introduce an optimization framework based on \textit{Magnitude Approximation} (MA) \cite{Kliuchnikov_2023}. Magnitude Approximation provides an approximating sequence of reduced $T$-count [$1 \log_2(1/\epsilon)$] as long as the approximation is allowed to have errors corresponding to residual rotations around {\it orthogonal} axes before and after the gate. By targeting a rotation with neighboring rotations on both sides within a quantum circuit for MA (e.g., the $Z$ rotation in an $X-Z-X$ sequence) and consolidating the residual error terms with the preexisting neighboring rotations, MA reduces the sequence $T$-count of the central rotation while maintaining approximation accuracy. Diagonal Approximation should then approximate the remaining rotations in this case.

However, applying MA effectively in a large-scale quantum circuit is non-trivial. The ability to absorb residual errors depends on the other gates in their neighborhood and on their commutation relations. Consequently, the decision to apply Magnitude versus Diagonal approximation for each rotation cannot be made locally and must take into consideration the decisions made for other rotations and the topology of the circuit; the optimal strategy involves a global delegation of approximation types across the entire circuit.

In this work, we address the combinatorial challenge of optimizing these decisions. We show that despite the configuration space scaling exponentially with circuit volume, the problem of selecting the optimal approximation strategy can be mapped to a classical 1D Ising model with a spatially varying field. By minimizing the energy of this system, we derive a linear-time algorithm (with respect to circuit volume) that identifies the globally optimal configuration. Benchmarking this approach against standard diagonal approximation on random quantum circuits, we observe an average reduction of 26\% in the total approximating circuit gate count.

\section{Preliminaries and Problem Definition}

\subsection{Standard Diagonal Approximation Baseline}
% For the specific and widely used Clifford+T gate set, more efficient number-theoretic approaches have been developed. The industry-standard baseline for synthesizing single-qubit gates is \textit{Diagonal Approximation} \cite{Kliuchnikov_2023}. 

\begin{figure}[htbp]
    \centering
    \includegraphics[width=0.8\linewidth]{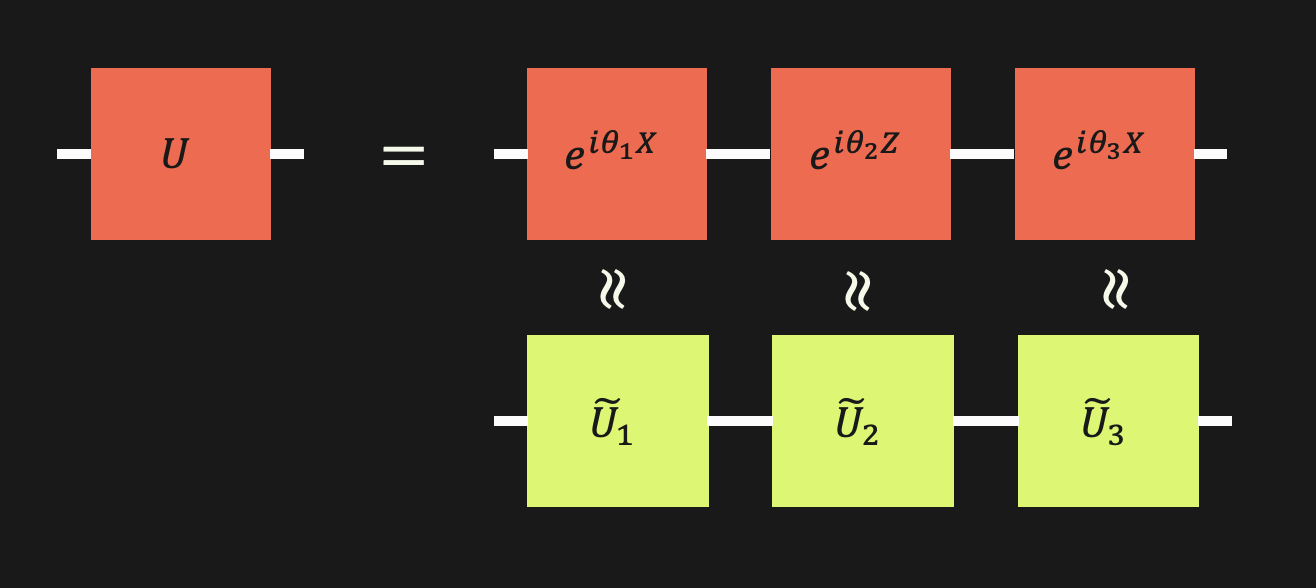}
    \caption{Decomposition of a single unitary operation $U$ into a sequence of three rotations ($X-Z-X$). In standard diagonal approximation, each rotation is approximated individually, resulting in a total T-count of approximately $9 \log_2(1/\epsilon)$.}
    \label{fig:euler_decomp}
\end{figure}

Since a general single-qubit unitary $U \in SU(2)$ is not a Pauli rotation (not ``diagonal''), it is typically decomposed into a sequence of three rotations using Euler angles (or Tait-Bryan angles), such as an $X-Z-X$ in \ref{fig:euler_decomp} sequence:
\begin{equation}
U = e^{i\theta_1 X} e^{i\theta_2 Z} e^{i\theta_3 X}
\end{equation}
In the standard approach, each of these three rotations is approximated individually using diagonal approximation. 

Because the approximation of a single diagonal rotation with precision $\epsilon$ requires a T-count of $N_T \approx 3\log_2(1/\epsilon)$, the total cost for synthesizing a general single-qubit unitary using this baseline method is:
\begin{equation}
N_T \approx 9 \log_2(1/\epsilon).
\end{equation}

More generally, to approximate an arbitrary multi-qubit circuit, each rotation gate requires an approximating sequence of $T$-count $N_T \approx 3\log_2(1/\epsilon)$. Note that the total accuracy budget for the approximation of the entire circuit should be partitioned between the various rotation gates such that in a circuit with more rotations, each rotation should be approximated with better accuracy.

\subsection{Magnitude Approximation (MA)}

\begin{figure}[htbp]
    \centering
    \includegraphics[width=1\linewidth]{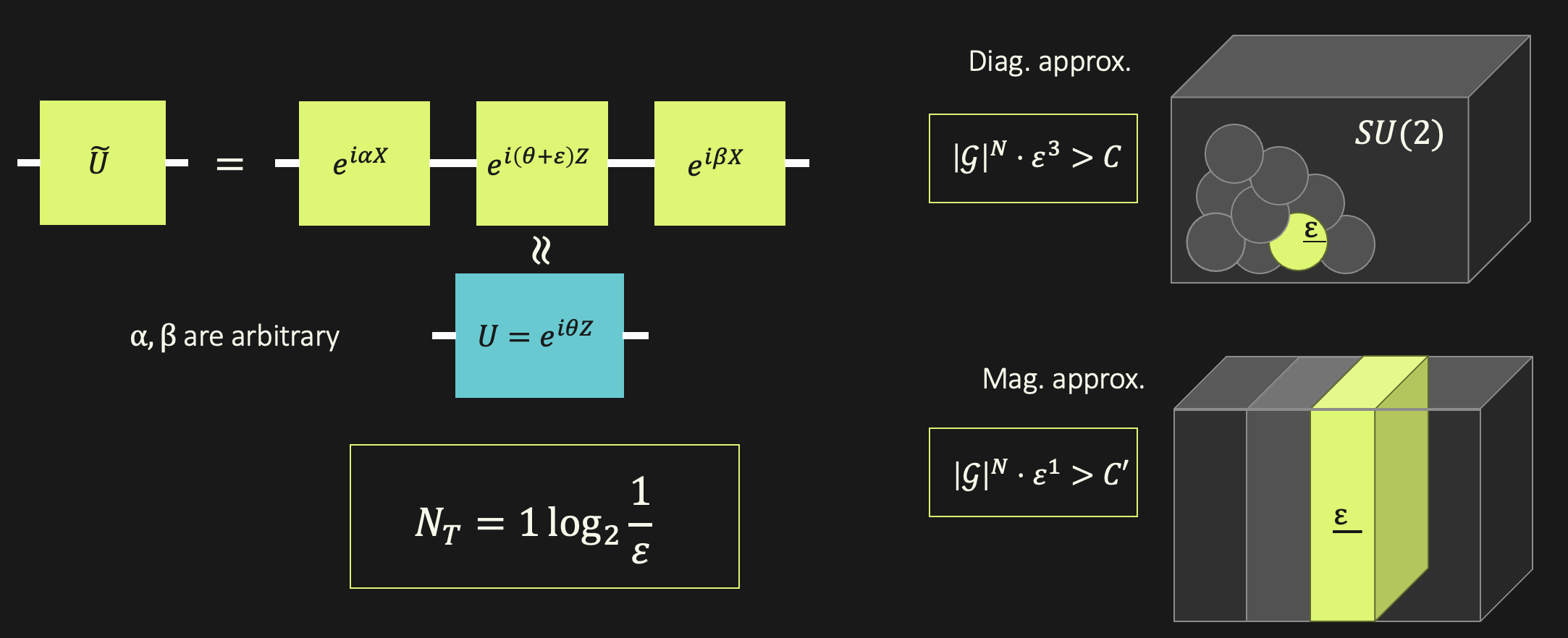}
    \caption{Concept of Magnitude Approximation. Left: A unitary is approximated by a central rotation with error $\epsilon$, while allowing arbitrary residual rotations $\alpha$ and $\beta$ on the orthogonal axis. These residuals are later absorbed into neighboring gates. Right: Geometrical comparison of the constraints imposed on the approximating unitary for Diagonal Approximation vs. Magnitude Approximation.}
    \label{fig:ma_concept}
\end{figure}

Given a target rotation (e.g., a Z-rotation $U \approx R_z(\theta)$), MA approximates it by a sequence $\tilde{U}$ as in Figure \ref{fig:ma_concept} such that:
\begin{equation}
\tilde{U} = e^{i\beta X} e^{i(\theta+\epsilon)Z} e^{i\alpha X}
\end{equation}
where $\epsilon$ is guaranteed to be smaller in magnitude than the permissible target, but $\alpha$ and $\beta$ are \textit{arbitrary} angles parameterizing residual rotations around an orthogonal axis (in this case represented as X rotations) \cite{Kliuchnikov_2023}. 

The key advantage of this relaxation is the significant reduction in T-gate cost. While fixing $\alpha$ and $\beta$ to specific values (as required in standard diagonal approximation) yields a sequence $T$-count of $N_T \approx 3 \log_2(1/\epsilon)$, leaving them arbitrary reduces the cost to:
\begin{equation}
N_T \approx 1 \log_2(1/\epsilon)
\end{equation}
This represents a $3\times$ reduction in the cost of the approximated gate itself, provided the residuals can be effectively managed.

The efficiency of MA relies on the ability to remove the arbitrary residuals $\alpha$ and $\beta$ from the circuit cost. This is achieved by absorbing them into neighboring gates. In a quantum circuit comprising rotations and CNOT gates, the residual rotations (e.g., $R_x(\alpha)$) can be propagated through the circuit using commutation relations until they merge with an adjacent rotation on the same axis or cancel out. The specific circuit transformations utilized include:

\begin{itemize}
    \item \textbf{Commutation:} $R_z$ gates commute with the control qubit of a CNOT, and $R_x$ gates commute with the target qubit. This allows residual rotations to "hop" over entangling gates to find a merger partner.
    \item \textbf{Merging:} Two adjacent rotations on the same axis, $R_x(\theta_1)$ and $R_x(\theta_2)$, can be merged into a single rotation $R_x(\theta_1 + \theta_2)$. If the residuals merge with existing rotations, they do not incur additional T-gate costs.
\end{itemize}

\section{Global Optimization via Ising Model Mapping}

\subsection{Circuit Segmentation and Canonical Forms}

To optimize the allocation of approximation strategies, we first map the quantum circuit to a discrete structure suitable for combinatorial optimization. As described in Ref.~\cite{iten2021introductionuniversalqcompiler}, quantum circuits containing CNOTs and single-qubit rotations can be reduced to a canonical form that includes a minimal number of rotations. Our optimization begins from this canonical form.

In our optimization framework, we treat the circuit as a sequence of segments on each qubit, separated by entangling gates that connect each qubit to any of the others.

\begin{enumerate}
    \item The circuit is divided into $N$ segments, where each segment $i$ consists of the single-qubit rotations situated between two consecutive CNOTs (or other Clifford gates) on the target qubit.
    \item In its canonical form, the circuit is pre-processed such that each segment contains a minimal set of rotations (up to 2 rotations per segment).
\end{enumerate}

As a key observation, we note that if Magnitude Approximation (MA) is chosen to be performed somewhere within a given segment, it is best first to combine all of the rotations associated with that segment (including those which may commute with the Clifford gates bounding the segment) into a generic single qubit unitary and decompose it as a sequence of three rotations with the axis of the first and last rotations chosen such that these may commute through the boundaries. This is always possible (there is always a third axis that can be selected for the central rotation, which differs from the other two). This procedure does not increase the number of rotations to be approximated since any Magnitude. 
Approximation within a segment necessarily results in a total of at least 3 rotations needing approximation within that segment (including the Magnitude Approximation residuals).

Next, we note that the best choice for where to perform Magnitude Approximation is on the central rotation in the new decomposition, since choosing one of the other two rotations would generate a new residual which would not commute with the boundaries of the segment and would therefore need to be approximated by itself. Furthermore, there is nothing to be gained by performing Magnitude Approximation more than once per segment. Consequently, the decision space to be optimized over reduces to a single binary decision per segment on each qubit. However, these decisions are not decoupled; the decisions in adjacent segments influence each other since rotations generated as magnitude approximation residuals may combine with preexisting rotations or with residual rotations in neighboring segments.

\subsection{The Hamiltonian Formulation}
We formulate the decision problem as a global optimization task mapped to a classical 1D Ising model with a spatially varying field.

Let $\sigma_i \in \{+1, -1\}$ be a binary decision variable associated with the $i$-th circuit segment:
\begin{itemize}
    \item $\sigma_i = +1$: Apply Magnitude Approximation.
    \item $\sigma_i = -1$: Don't apply Magnitude Approximation.
\end{itemize}

The objective is to minimize the total T-gate count required to approximate the circuit. This cost function is expressed as a Hamiltonian $H$ (up to additive constants):
\begin{equation}
H(\sigma) = -\sum_{i} \left( J_{i+\frac{1}{2}} \sigma_i \sigma_{i+1} + h_i \sigma_i \right)
\end{equation}
\label{eq: hamiltonian}
Here, $h_i$ represents the local field (biasing the decision based on the intrinsic benefit of MA for segment $i$) and $J_{i+\frac{1}{2}}$ represents the interaction term (capturing the cost or penalty of mismatching strategies between neighbors).

\begin{figure}[htbp]
    \centering
    \includegraphics[width=1\linewidth]{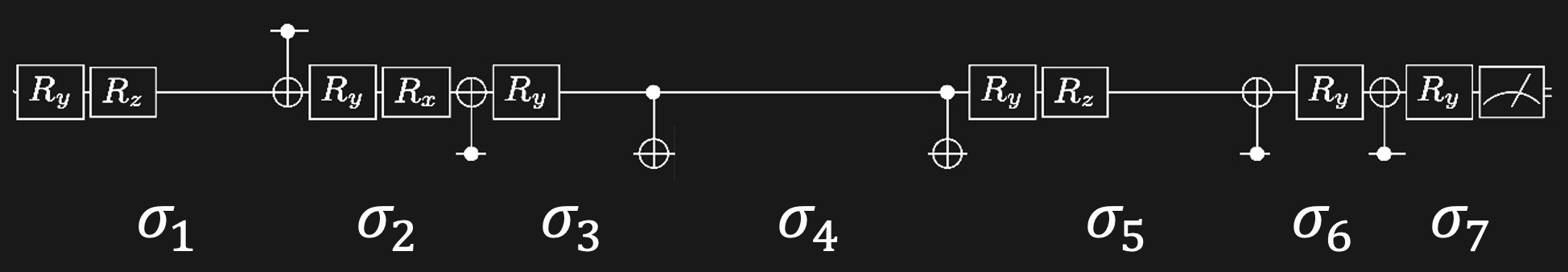}
    \caption{Mapping the quantum circuit optimization problem to a 1D Ising model. Each circuit segment between CNOTs is treated as a site with spin $\sigma_i$, where the interaction $J$ and field $h$ are determined by the number of rotations $r_i$ confined to each segment and the number of rotations which can move through segment boundaries $r_{i+\frac{1}{2}}$.}
    \label{fig:ising_mapping}
\end{figure}

The specific values of the coefficients $h_i$ and $J_{i+\frac{1}{2}}$ are derived from the trade-off between the gate reduction offered by MA for the targeted query rotation and the need to approximate then the residuals created through this process:

\begin{equation}
h_i = \left(\frac{3}{2}(r_i+r_{i+\frac{1}{2}})-2\right)\log_2(1/\epsilon),
\end{equation}
\label{eq: h}

\begin{equation}
J_{i+\frac{1}{2}} = \frac{3}{4}\left(1-r_{i+\frac{1}{2}}\right)\log_2(1/\epsilon),
\end{equation}
\label{eq: J}

Where the coefficients are functions of the number of rotations gates confined within each segment $r_i$ and the number of rotation gates that commute with each boundary between two segments $r_{i+\frac{1}{2}}$. In essence, Magnitude Approximation is beneficial in regions with many rotations and when neighboring segments employ Magnitude Approximation as well.

\section{Linear-Time Optimization Algorithm}

While the configuration space of the approximation strategies scales exponentially with the circuit volume ($2^N$ for $N$ segments), the mapping to a 1D Ising model allows for an exact optimal solution in linear time. We employ a recursive dynamic programming algorithm that identifies the ground state of the Hamiltonian defined in the previous section \ref{eq: hamiltonian}.

\subsection{Recursive Energy Minimization (Forward Pass)}

The algorithm proceeds by calculating the minimum cumulative energy to reach a specific state ($\sigma_i = +1$ or $\sigma_i = -1$) at segment $i$, based on the optimal path to segment $i-1$.

Let $E^+_i$ be the minimum energy of the partial chain ending at segment $i$ with $\sigma_i = +1$.
Let $E^-_i$ be the minimum energy of the partial chain ending at segment $i$ with $\sigma_i = -1$.

For the first segment ($i=1$), the energies are initialized based solely on the local field:
\begin{equation}
E^+_1 = -h_1, \quad E^-_1 = h_1
\end{equation}

For each subsequent segment $i$ from 2 to $N$, we compute the optimal energy by considering both possible states of the previous neighbor ($i-1$) and the interaction term $J_{i-1}$. For example, to find $E^+_i$, we compare the cost of transitioning from $\sigma_{i-1}=+1$ versus $\sigma_{i-1}=-1$:
\begin{equation}
E^+_i = \min \left( E^+_{i-1} - J_{i-1} - h_i, \quad E^-_{i-1} + J_{i-1} - h_i \right)
\end{equation}
A similar minimization is performed for $E^-_i$. During this forward pass, we store a pointer to the previous state that yielded the minimum energy.

\begin{algorithm}[H]
\caption{Linear-Time Recursive Solver for Approximation Strategy}
\begin{algorithmic}[5]
\State \textbf{Input:} Couplings $J[1\dots N-1]$, Fields $h[1\dots N]$
\State \textbf{Initialize:} 
\State $E_{plus}[1] \gets -h[1]$
\State $E_{minus}[1] \gets h[1]$
\For{$i = 2$ to $N$}
    \State \textit{// Compute min energy ending in MA (+1)}
    \If{$E_{plus}[i-1] - J[i-1] < E_{minus}[i-1] + J[i-1]$}
        \State $E_{plus}[i] \gets E_{plus}[i-1] - J[i-1] - h[i]$
        \State $trace[i][+1] \gets +1$
    \Else
        \State $E_{plus}[i] \gets E_{minus}[i-1] + J[i-1] - h[i]$
        \State $trace[i][+1] \gets -1$
    \EndIf
    
    \State \textit{// Compute min energy ending in Diagonal (-1)}
    \If{$E_{plus}[i-1] + J[i-1] < E_{minus}[i-1] - J[i-1]$}
        \State $E_{minus}[i] \gets E_{plus}[i-1] + J[i-1] + h[i]$
        \State $trace[i][-1] \gets +1$
    \Else
        \State $E_{minus}[i] \gets E_{minus}[i-1] - J[i-1] + h[i]$
        \State $trace[i][-1] \gets -1$
    \EndIf
\EndFor
\State \Return $E_{plus}, E_{minus}, trace$
\end{algorithmic}
\label{alg:recursive_optimization}
\end{algorithm}

\subsection{Backtracking for Optimal Configuration}
Once the forward pass is complete, the global minimum energy is determined by comparing the final states: $\min(E^+_N, E^-_N)$. The optimal configuration $\vec{\sigma}$ is then reconstructed via backtracking.

We set the state of the last segment, $\sigma_N$, to the lower-energy state. We then iteratively determine the state of the preceding segment $\sigma_{i-1}$ using the stored `spin\_trace` values:
\begin{equation}
\sigma_{i-1} = \text{trace}[i][\sigma_i]
\end{equation}
This process continues down to $i=1$. This algorithm solves the optimization problem with time complexity $O(N)$, where $N$ is the number of segments, making it scalable for large quantum circuits where brute-force methods are infeasible.

\section{Benchmarking and Results}

To validate the efficiency of the proposed global optimization framework, we benchmark the algorithm against the strategy of using only Diagonal Approximation on randomized quantum circuits.

The benchmarks were conducted on a set of random quantum circuits consisting of single-qubit rotations interspersed with CNOTs. These circuits represent typical ``unstructured'' quantum algorithms or random circuit sampling tasks, which serve as a robust stress test for compilation strategies. We compared the total T-gate count of the approximating circuits generated by our global optimization algorithm (finding the ground state of the Ising Hamiltonian) against the baseline. Across the tested benchmark set, the global optimization achieved an average reduction of \textbf{26\%} in the total approximating circuit T-gate count. We further test our optimization protocol on a suite of molecular simulation circuits generated by the Classiq platform \cite{classiq}. The results achieved on these circuits are highlighted in Table.~\ref{tab:results}. The reduction achieved in the T-gate count varies between circuits for different molecules, ranging between 13\% and 58\%.

\begin{table}[htbp]
\centering
\caption{Comparison of T-gate counts between the standard Diagonal Approximation baseline and the proposed Global Optimization (Magnitude Approximation) for various molecular simulation circuits. The reduction percentage highlights the efficiency gain achieved through our optimization protocol.}
\vspace{0.2cm}
\begin{tabular}{l c c c}
\hline
\textbf{Circuit /} & \textbf{Baseline} & \textbf{Optimized} & \textbf{Reduction (\%)} \\
\textbf{Molecule} & \textbf{T-Count} & \textbf{T-Count} &  \\
\hline
$H_2$ & 534 & 464 & 13\% \\
$H_3$ & 2,015 & 1,446 & 28\% \\
$H_4$ & 9,452 & 6,483 & 31\% \\
$H_2O$ & 11,576 & 8,123 & 29\% \\
$H_5$ & 23,235 & 15,751 & 32\% \\
UCCSD & 28,493 & 11,883 & 58\% \\
$H_2O$ (large) & 43,833 & 30,166 & 31\% \\
\hline
\end{tabular}
\label{tab:results}
\end{table}

\section{Conclusion}

We have introduced a linear-time optimization algorithm for determining the strategy used for each single-qubit rotation synthesis within a larger-scale quantum circuit. The algorithm maps the circuit configuration to a classical 1D Ising model. By defining a Hamiltonian that captures the trade-off between the local efficiency of Magnitude Approximation and the global constraints of residual error absorption, we replace heuristic greedy methods with a globally optimal solution. As quantum compilers evolve to handle larger and more complex algorithms \cite{classiq}, such global optimization strategies will be essential for maximizing the computational power of limited quantum hardware.

\bibliographystyle{plain}
\bibliography{references}

\end{document}